\documentclass[prb,aps,twocolumn,floatfix,superscriptaddress]{revtex4-2}
\usepackage{float}
\usepackage{color}
\usepackage{amsmath}
\usepackage{latexsym}
\usepackage{amssymb}
\usepackage{amsmath}
\usepackage{amsfonts}
\usepackage{mathrsfs}
\usepackage{graphics,epstopdf}
\usepackage{placeins}
\usepackage[colorlinks=true, citecolor=blue, urlcolor=blue, linkcolor=blue ]{hyperref}
\usepackage{epsf,graphics,graphicx}

\date{\today}
\usepackage{bigints}
\usepackage{multirow}
\usepackage{braket}
\usepackage{makecell}
\begin{document}

\title
  {Strong Optical–Optical Avoided Crossings Suppress Thermal Conductivity in Ga-Substituted TlInTe$_2$}

\author{Sayan Paul}
\author{Swapan K. Pati}
\email{pati@jncasr.ac.in}
\affiliation
{Theoretical Sciences Unit, School of Advanced Materials (SAMat), Jawaharlal Nehru Centre for Advanced Scientific Research, Bangalore 560064, India}
\date{\today} 

\begin{abstract}
In crystalline solids, avoided crossing between acoustic and optical phonons is widely recognized as an effective mechanism for suppressing lattice thermal conductivity ($\kappa_l$). However, the role of avoided crossings among optical phonons remains largely unexplored due to their weak contribution to heat transport. Here, using first-principles calculations combined with the linearized Wigner transport equation (LWTE), we demonstrate that optical–optical avoided crossings can effectively reduce ($\kappa_l$) in TlIn$_{0.5}$Ga$_{0.5}$Te$_2$. Pristine TlInTe$_2$ exhibits strong optical phonon-dominated heat transport, where optical phonons contribute nearly 63\%\ of $\kappa_l$. The phonon dispersion of TlInTe$_2$ shows several crossing points in the optical region, which evolve into avoided crossings after 50\%\ Ga substitution. Irreducible representation analysis reveals that the crossing phonon branches in TlInTe$_2$ belong to different symmetry representations, whereas the corresponding branches in TlIn$_{0.5}$Ga$_{0.5}$Te$_2$ possess the same symmetry representation, which enables phonon modes to couple and results in gap opening at the crossing points. These avoided crossings significantly suppress the optical phonon group velocity, thereby reducing the optical phonon contribution from 63\%\ to 44\%\ and lowering $\kappa_l$ from 0.568 to 0.482 Wm$^{-1}$K$^{-1}$ at 300 K. Mode-averaged transport analysis further confirms that the suppression of $\kappa_l$ is primarily governed by reduced phonon group velocity ($v_g$), while enhanced anharmonic scattering provides an additional secondary contribution. Our results establish symmetry-modified optical–optical avoided crossing as an effective route to suppress optical phonon transport and reduce $\kappa_l$ in systems where optical phonons significantly contribute to heat transport.  
\end{abstract}

\maketitle

\section{Introduction}
Materials exhibiting intrinsically low lattice thermal conductivity ($\kappa_l$) are of significant interest for their application in thermoelectrics,\cite{tan2016rationally, qian2021phonon, sarkar2024hidden} thermal barrier coatings,\cite{padture2002thermal} heat management, \cite{HUANG2023120745} and photovoltaics, \cite{green2017energy} etc. Thus, understanding and engineering phonon dynamics and phonon-mediated thermal transport to achieve reduced $\kappa_l$ are extremely crucial. Over the years, various intrinsic and extrinsic approaches have been explored to tailor lattice thermal transport. Intrinsic strategies, such as bonding heterogeneity,\cite{pal2019high} rattler-like atoms,\cite{christensen2008avoided, lin2016concerted} lattice instability associated with Pauling's 3rd rule,\cite{pathak2024deciphering, pathak2026anharmonicity} liquid-like sublattice,\cite{liu2012copper} ferroelectric instability,\cite{zhang2011anomalous} antibonding states below the Fermi level,\cite{he2022accelerated} etc. are widely used to design novel low $\kappa_l$ materials. On the other hand, extrinsic approaches, including all-scale hierarchical phonon engineering,\cite{biswas2012high, pei2017integrating} entropy-driven alloying,\cite{jiang2021high, bhui2025atomic} defect engineering,\cite{zheng2021defect} etc. are employed to modulate thermal transport of a material.

Within the phonon gas model, the lattice thermal conductivity of crystalline solids is expressed as $\kappa_l=\frac{1}{3}C_V v_g l$, where $C_V$ is the phonon specific heat, $v_g$ is the group velocity, and $l$ is the phonon mean free path.\cite{beekman2017inorganic} So, the acoustic phonon modes, which are typically highly dispersive and possess large $v_g$ and long $l$, dominate in heat transport. In contrast, optical phonons are generally less dispersive, leading to lower $v_g$ and a short mean free path. Therefore, their contribution to heat transport is significantly less compared to acoustic phonons. Thus, controlling the dispersion of acoustic phonons is one of the primary strategies to suppress the $\kappa_l$.\cite{duhan2025strong, li2016influence} In several halide and chalcogenide-based compounds, such as Cs$_3$Bi$_2$I$_6$Cl$_3$,\cite{acharyya2022glassy, zeng2025lattice} KCu$_5$Se$_3$,\cite{li2023overdamped} TlCuSe,\cite{lin2021ultralow} and Tl$_2$AgI$_3$\cite{pathak2026anharmonicity} etc., the presence of low-frequency localized optical phonon modes has been shown to strongly scatter heat-carrying acoustic phonons, which reduces $\kappa_l$. In cage-like materials such as skutterudites and clathrates, the loosely bound guest atoms move independently from the host matrix, which is referred as ``rattling" motion or Einstein-like motion.\cite{sales1996filled, rull2015skutterudites, takabatake2014phonon} These rattling vibrations give rise to low-frequency, localized optical phonon modes that lie in close proximity to acoustic branches.\cite{dong2001theoretical} Inelastic neutron scattering experiments have shown that these modes hybridize with acoustic phonons, leading to avoided crossings that suppress the dispersion of acoustic branches, reduce their $v_g$ and effectively suppress $\kappa_l$.\cite{christensen2008avoided} 

\begin{figure}[t] 
  \centering
  \includegraphics[width=\columnwidth]{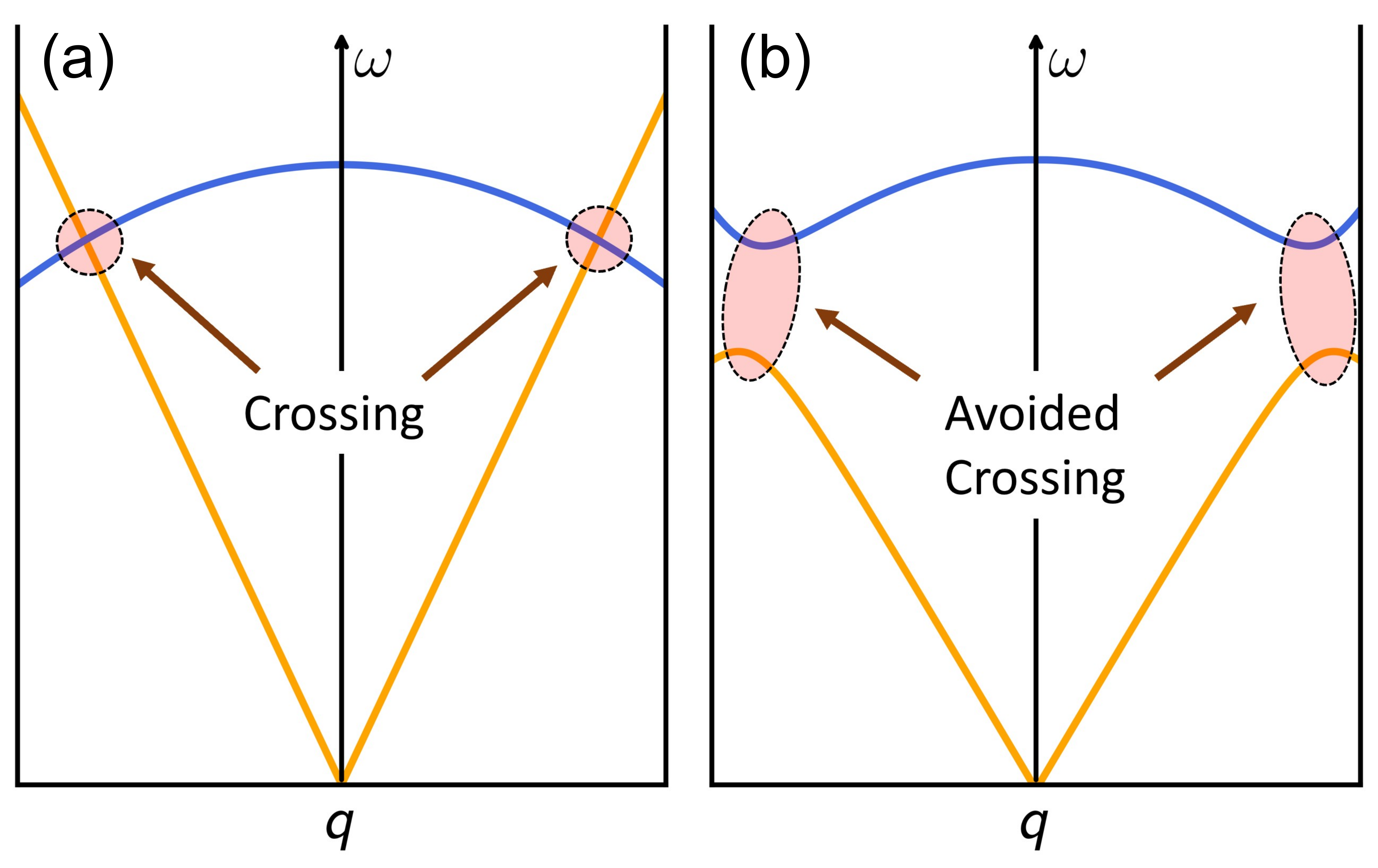}
  \caption{Schematic illustration of (a) phonon branch crossing and (b) avoided crossing arising from mode coupling. In (a), two non-interacting phonon modes intersect due to symmetry incompatibility. In (b), interaction between modes with compatible symmetry leads to hybridization, resulting in band repulsion, and a finite frequency gap at the crossing point leads to the avoided crossing region.}
  \label{fig:scheme}
\end{figure}

Avoided crossing in phonon arises when two phonon modes of same symmetry interact and hybridize with each other, so they do not intersect.\cite{christensen2008avoided} Instead of crossing, they repel each other and a characteristic gap opening is observed at the crossing point, which reduces their group velocity (FIG.~\ref{fig:scheme}). Avoided crossings between acoustic and optical phonons have been extensively studied as an effective strategy to reduce $\kappa_l$ in many compounds.\cite{han2023strong, li2015ultralow} However, avoided crossing between optical phonons has received very little attention due to the small contribution of optical phonons to $\kappa_l$. Interestingly, recent studies have revealed that optical phonons can contribute significantly to $\kappa_l$ in certain systems. For example, in SnSe\cite{PhysRevB.92.115202} and BaSnS$_2$\cite{li2021optical} crystals, $\sim$60\% and $\sim$68\% of total $\kappa_l$ is contributed by the optical phonons, respectively. In such cases, engineering the dispersion of optical phonons provides an alternative pathway to modulate thermal transport.

In this work, we demonstrate that chemical substitution modifies the symmetry characteristics of optical phonon modes, leading to avoided crossings among optical phonons and reducing the overall $\kappa_l$. We choose a well-studied ternary chalcogenide compound TlInTe$_2$ from the ABX$_2$ (A = In$^+$, Tl$^+$, B = Ga$^{3+}$, In$^{3+}$, Tl$^{3+}$, X = Se$^{2-}$, Te$^{21}$) family,  which shows ultra-low $\kappa_l$ due to the presence of rattler cation at A site.\cite{jana2017intrinsic, pal2021microscopic} We observe several crossings in the phonon dispersion at the optical phonon region of pristine TlInTe$_2$. Due to the high dispersion of these modes, we show that optical phonons contribute $\sim$63\% to heat transport, as evidenced by cumulative $\kappa_l$ analysis. Upon 50\% Ga substitution at the In site, the crossing points evolve into avoided crossings, resulting in pronounced flattening of the optical branches and a corresponding reduction in $v_g$. Consequently, optical phonon contribution to $\kappa_l$ is reduced to $\sim$44\%, which suppress the overall $\kappa_l$ of TlIn$_{0.5}$Ga$_{0.5}$Te$_2$ to 0.482 Wm$^{-1}$K$^{-1}$ from 0.568 Wm$^{-1}$K$^{-1}$ of TlInTe$_2$. Thus, our results establish a distinct mechanism of modulating $\kappa_l$ by tuning of the optical phonons effectively.

\section{Computational Methodology}
We carried out first-principles DFT calculations in the Vienna Ab initio Simulation Package (VASP)\cite{kresse1996efficiency, kresse1996efficient} using the projected augmented wave (PAW)\cite{kresse1999ultrasoft, blochl1994projector} pseudopotentials of Tl (5d$^{10}$ 6s$^2$, 6p$^1$), In (4d$^{10}$ 5s$^2$, 5p$^1$), Ga (3d$^{10}$ 4s$^2$, 4p$^1$) and Te (5s$^2$, 5p$^4$). The Perdew, Burke and Ernzerhof functional for solids (PBEsol)\cite{perdew1996generalized, perdew2008restoring} within the generalized gradient approximation (GGA) is used for describing the exchange-correlation energies. The electronic wavefunctions were expanded using a plane wave basis set with a 650 eV kinetic energy cutoff, and a $\Gamma$-centered 12$\times$12$\times$12 mesh was employed for the Brillouin zone sampling. The threshold criteria for energy and Hellmann-Feynman force convergence were set to 10$^{-8}$
eV and 10$^{-4}$ eV/\AA, respectively.

The harmonic phonon dispersion was calculated using finite displacement method as implemented in the Phonopy package.\cite{togo2023first, togo2023implementation} We took a 3$\times$3$\times$3 supercell (216 atoms) of the primitive unit cell (8 atoms) to generate the symmetry-inequivalent displaced structures, from which the dynamical matrix and harmonic second-order interatomic force constants (IFCs) were constructed via static DFT calculations.

To obtain the phonon relaxation time ($\tau$) and anharmonic phonon properties, we constructed 3rd order IFC matrix from the displaced configuration generated from a 2$\times$2$\times$2 supercell (64 atoms) using the Phono3py package.\cite{togo2023first, togo2023implementation, togo2015distributions} Then, the $\kappa_l$ and related anharmonic transport properties were evaluated by solving the linearized Wigner transport equation (LWTE),\cite{simoncelli2019unified, simoncelli2022wigner} where $\kappa_l$ is expressed as
\begin{equation}
    \kappa_l^{\alpha\beta}=\kappa_p^{\alpha\beta}+\kappa_c^{\alpha\beta}.
    \label{eq:kl}
\end{equation}
Here, $\kappa_p$ and $\kappa_c$ represent the population (particle-like) and coherence (wave-like) contributions to $\kappa_l$, respectively. In $\kappa_p$ mechanism heat is transported by the diagonal elements of the heat flux operator, which is same as the Boltzmann transport equation (BTE). $\kappa_p$ is calculated by the direct solution of the BTE using the equation\cite{chaput2013direct}
\begin{equation}
    \kappa_{p}^{\alpha\beta}
    = \frac{\hbar^{2}}{4 k_{B} T^{2} N V_{0}}
      \sum_{\lambda \lambda'}
      \frac{\omega_{\lambda} v^{\alpha}(\lambda)}{\sinh\!\left( \frac{\hbar \omega_{\lambda}}{2 k_{B} T} \right)}
      \frac{\omega_{\lambda'} v^{\beta}(\lambda')}{\sinh\!\left( \frac{\hbar \omega_{\lambda'}}{2 k_{B} T} \right)}
      \left( \Omega^{\sim1} \right)_{\lambda \lambda'},
    \label{eq:kp}
\end{equation}
where $\Omega^{\sim1}$ denotes the Moore-Penrose inverse of the collision matrix $\Omega$. The collision matrix is

\begin{eqnarray}
    \Omega_{\lambda \lambda'} = \delta_{\lambda \lambda'} \frac{1}{\tau_{\lambda}} 
    + \frac{\pi}{\hbar^{2}} \sum_{\lambda''} 
    \left| \Phi_{\lambda \lambda' \lambda''} \right|^{2} 
    \frac{1}{\sinh \left( \frac{\hbar \omega_{\lambda''}}{2 k_{B} T} \right)} \\ \nonumber
    \times \left[ \delta(\omega_{\lambda} - \omega'_{\lambda} - \omega''_{\lambda}) 
    + \delta(\omega_{\lambda} + \omega'_{\lambda} - \omega''_{\lambda}) 
    + \delta(\omega_{\lambda} - \omega'_{\lambda} + \omega''_{\lambda}) \right].
\end{eqnarray}
\begin{figure*}[t] 
  \centering
  \includegraphics[width=1\linewidth]{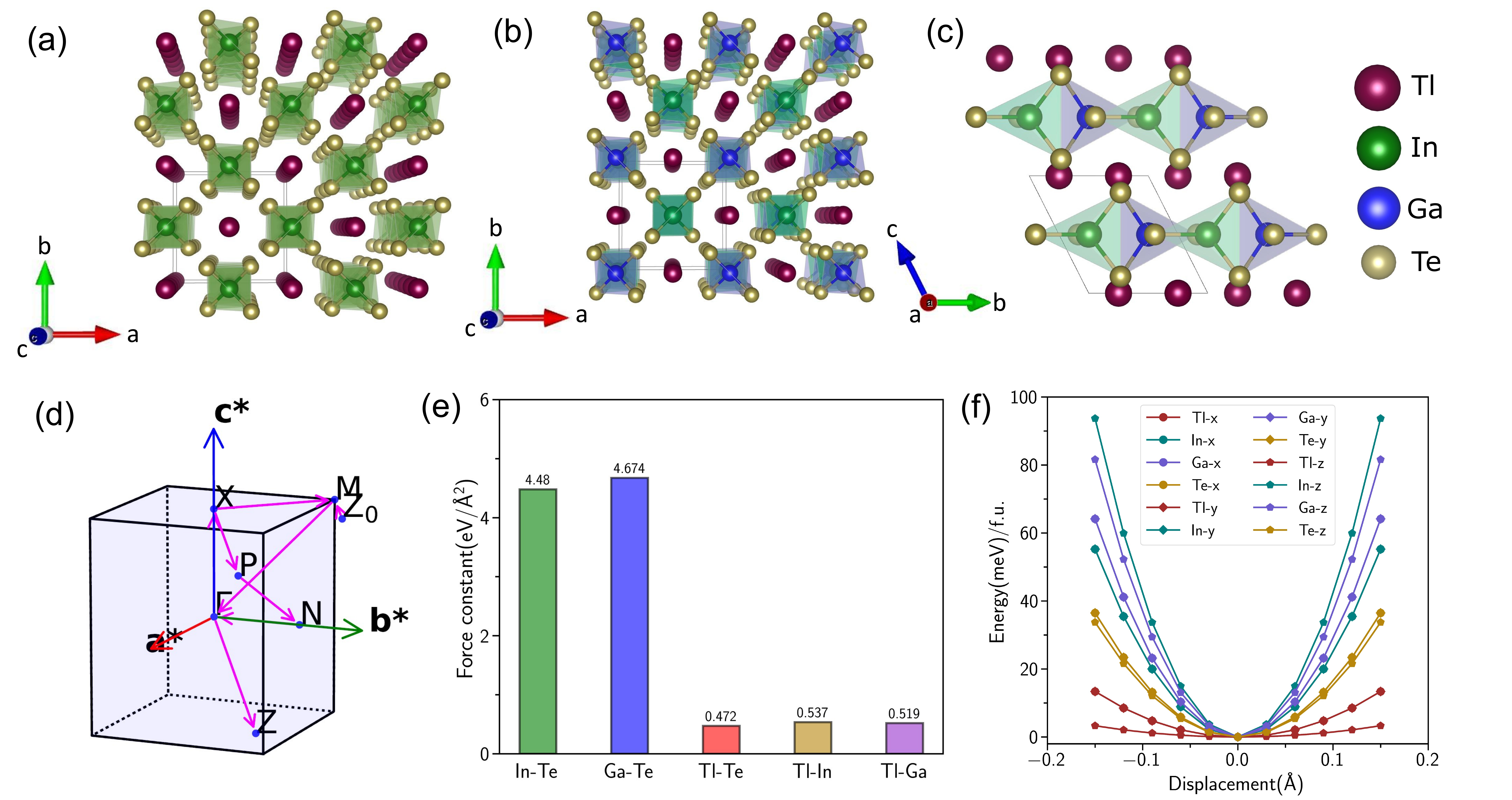}
  \caption{Conventional crystal structure of (a) TlInTe$_2$ and (b) TlIn$_{0.5}$Ga$_{0.5}$Te$_2$, viewed along crystallographic $c$-axis. The crystal structures show a one-dimensional tetrahedral chain extended along the $c$-axis.(c) Primitive crystal structure of TlIn$_{0.5}$Ga$_{0.5}$Te$_2$, viewed along crystallographic $a$-axis. (d) The first Brillouin zone of these compounds in the $k$-space. (e) Second order interatomic force constants (IFC) between the atom pairs. (f) The potential energy surface (PES) of all atoms along $x$, $y$, and $z$-directions.}
  \label{fig:structure}
\end{figure*}
Here, $k_B$ is the Boltzmann constant, $\omega_\lambda$ is the frequency of the phonon of mode $\lambda$, $v^{\alpha}(\lambda)$ is the velocity of $\lambda$ phonon mode in $\alpha$ direction, T is temperature, $\Phi_{\lambda \lambda' \lambda''}$ is 3rd order IFC and $\tau_{\lambda}$ is phonon relaxation time. The equation is solved with a 12×12×12 $q$-point mesh employing 2nd and 3rd order IFCs. The convergence of $\kappa_l$ with $q$-mesh is shown in FIG.~\textcolor{blue}{S2, Supplemental Material}. Conversely, the coherence term $\kappa_c$ originates from the off-diagonal elements of the heat flux operator, which captures wavelike heat transport arising from coupling between quasi-degenerate phonon modes. This mechanism is analogous to Allen-Feldmann theory for disordered solids.\cite{allen1989thermal, allen1993thermal} In the LWTE formalism $\kappa_c$ is calculated using the equation\cite{simoncelli2019unified, simoncelli2022wigner} 
\begin{eqnarray}
    \kappa_c^{\alpha \beta} = \frac{\hbar^2}{k_B T^2} \frac{1}{\mathcal{V}N_c} \sum_{\mathbf q} \sum_{s \neq s'} \frac{\omega(\mathbf q)_s + \omega(\mathbf q)_{s'}}{2} v^\alpha(\mathbf q)_{s,s'} v^\beta(\mathbf q)_{s',s} \\ \nonumber  
    \times \frac{\omega(\mathbf q)_s \bar{N}^T(\mathbf q)_s [\bar{N}^T(\mathbf q)_s + 1] + \omega(\mathbf q)_{s'} \bar{N}^T(\mathbf q)_{s'} [\bar{N}^T(\mathbf q)_{s'} + 1]}{4[\omega(\mathbf q)_{s'} - \omega(\mathbf q)_s]^2 + [\Gamma(\mathbf q)_s + \Gamma(\mathbf q)_{s'}]^2} \\ \nonumber
    \times [\Gamma(\mathbf q)_s + \Gamma(\mathbf q)_{s'}]
    \label{eq:kc}
\end{eqnarray}
where $\mathcal{V}$ is the volume of the primitive cell, $N_c$ is the number of phonon eigenvectors, $\textbf{q}$ and $s$ are wavevector and branch index respectively, $\bar{N}^T(q)_s$ is the Bose-Einstein distribution at temperature T, and $\Gamma(\textbf{q})_s$ is the phonon linewidth, which is inverse proportional to the phonon lifetime ($\Gamma(\textbf{q})_s=\Gamma_{\lambda}=1/{\tau_{\lambda}}$).

\section{Results and Discussion}

\begin{figure}[t] 
  \centering
  \includegraphics[width=0.8\columnwidth]{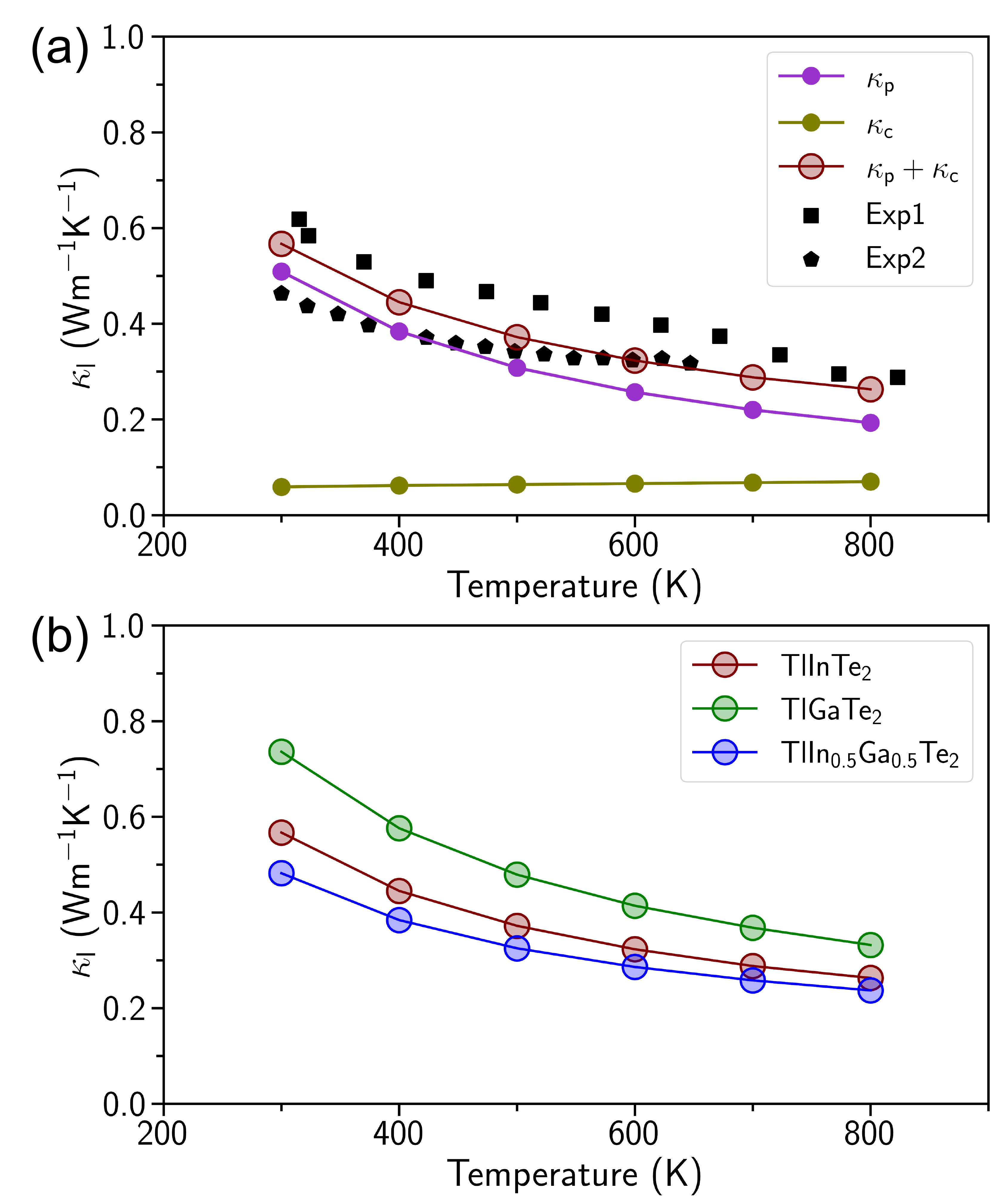}
  \caption{(a) Temperature dependence of the calculated average population conductivity ($\kappa_p$), coherence conductivity ($\kappa_c$), and their summation ($\kappa_l=\kappa_p+\kappa_c$) of TlInTe$_2$ using LWTE and compared with previously experimentally reported $\kappa_l$ data from Exp1\cite{jana2017intrinsic} and Exp2\cite{matsumoto2008systematic}. (b) Temperature-dependent $\kappa_l$ of TlIn$_{0.5}$Ga$_{0.5}$Te$_2$ and compared with TlInTe$_2$ and TlGaTe$_2$.}
  \label{fig:kl}
\end{figure}

TlInTe$_2$ crystallizes in a tetragonal body-centered crystal structure with a space group of I4/mcm (FIG.~\ref{fig:structure}a). In this compound Tl, In and Te atoms adopt the oxidation states of +1, +3, and -2, respectively. The structural framework is built from In$^{3+}$ cations tetrahedrally coordinated by Te$^{2-}$ anions, which form [InTe$_4$]$^-$ covalently bonded anionic sublattice. These [InTe$_4$]$^-$ tetrahedra share edges with neighbouring units and give rise to one-dimensional chains extending along the crystallographic \textit{c}-axis, but they are weakly connected in the crystallographic \textit{ab}-plane, with no direct covalent bonding between adjacent chains. The Tl$^+$ ions occupy the interstitial regions between these chains and interact ionically with the [InTe$_4$]$^-$ framework.\cite{jana2017intrinsic, pal2021microscopic}

FIG.~\ref{fig:structure}b and \ref{fig:structure}c represent the conventional and primitive structures of TlIn$_{0.5}$Ga$_{0.5}$Te$_2$, respectively, where one In atom is replaced by Ga atom. In this structure, each one-dimensional chain consists of InTe$_4$ and GaTe$_4$ tetrahedra alternately (FIG.~\ref{fig:structure}c). The optimized lattice parameters of TlInTe$_2$ are \textit{a}=\textit{b}=8.41 \AA\ and \textit{c}=7.14 \AA, which are in good agreement with the experimental values of \textit{a}=\textit{b}=8.48 \AA\ and \textit{c}=7.19 \AA. In TlIn$_{0.5}$Ga$_{0.5}$Te$_2$, the lattice parameters are reduced to \textit{a}=\textit{b}=8.38 \AA\ and \textit{c}=6.96 \AA\, which is attributed to the smaller size of Ga compared to In atoms. Both compounds exhibit semiconducting behaviour, as evidenced by the electronic band structures shown in FIG. \textcolor{blue}{S1, Supplemental Material}.

To understand the bonding environment of TlIn$_{0.5}$Ga$_{0.5}$Te$_2$, we analyzed the second order interatomic force constants (IFC) and potential energy surface (PES) of all atoms (FIG.~\ref{fig:structure}e and \ref{fig:structure}f). As shown in Fig.~\ref{fig:structure}e, the IFC values for In–Te and Ga–Te bonds are 4.48 and 4.67 eV/\AA$^2$, respectively, which are significantly larger than those involving Tl (Tl–Te, Tl–In, and Tl–Ga). This indicates In-Te and Ga-Te bonds are strong and towards covalent in nature, while Tl atoms interact weakly with the surrounding [InTe$_4$]$^-$ and [GaTe$_4$]$^-$ anionic sublattices, consistent with ionic interactions. Further, we observe the IFC between Ga-Te is slightly higher compared to In-Te, which is due to the shorter distance between Ga and Te atoms (2.67 \AA) than the In and Te atoms (2.81 \AA). The potential energy surface profiles in Fig.~\ref{fig:structure}f further support this picture. We observe that In and Ga have a deep PES and Tl has a shallow PES, indicating In and Ga are very strongly bonded with the lattice, but Tl atoms are very loosely bonded with the lattice. This confirms that Tl atoms can vibrate with a large amplitude in the lattice, confirming the rattling nature of Tl atoms. On a closer examination, we observe that the PES of In and Ga along the $z$-axis is slightly deeper than $x$- and $y$-axes. This anisotropy arises from the strong one-dimensional tetrahedral chains extending along the crystallographic $c$-axis, whereas interactions along the $a$- and $b$-axes are comparatively weaker. The PES of Te atoms lies intermediate between those of In/Ga and Tl, indicating that Te atoms are less strongly bound within the one-dimensional chain compared to In and Ga, but more strongly bound than the loosely confined Tl atoms.  From the anisotropy in the PES, we can expect a similar anisotropy in the lattice thermal conductivity ($\kappa_l$) in TlIn$_{0.5}$Ga$_{0.5}$Te$_2$.

\begin{figure}[t] 
  \centering
  \includegraphics[width=1\columnwidth]{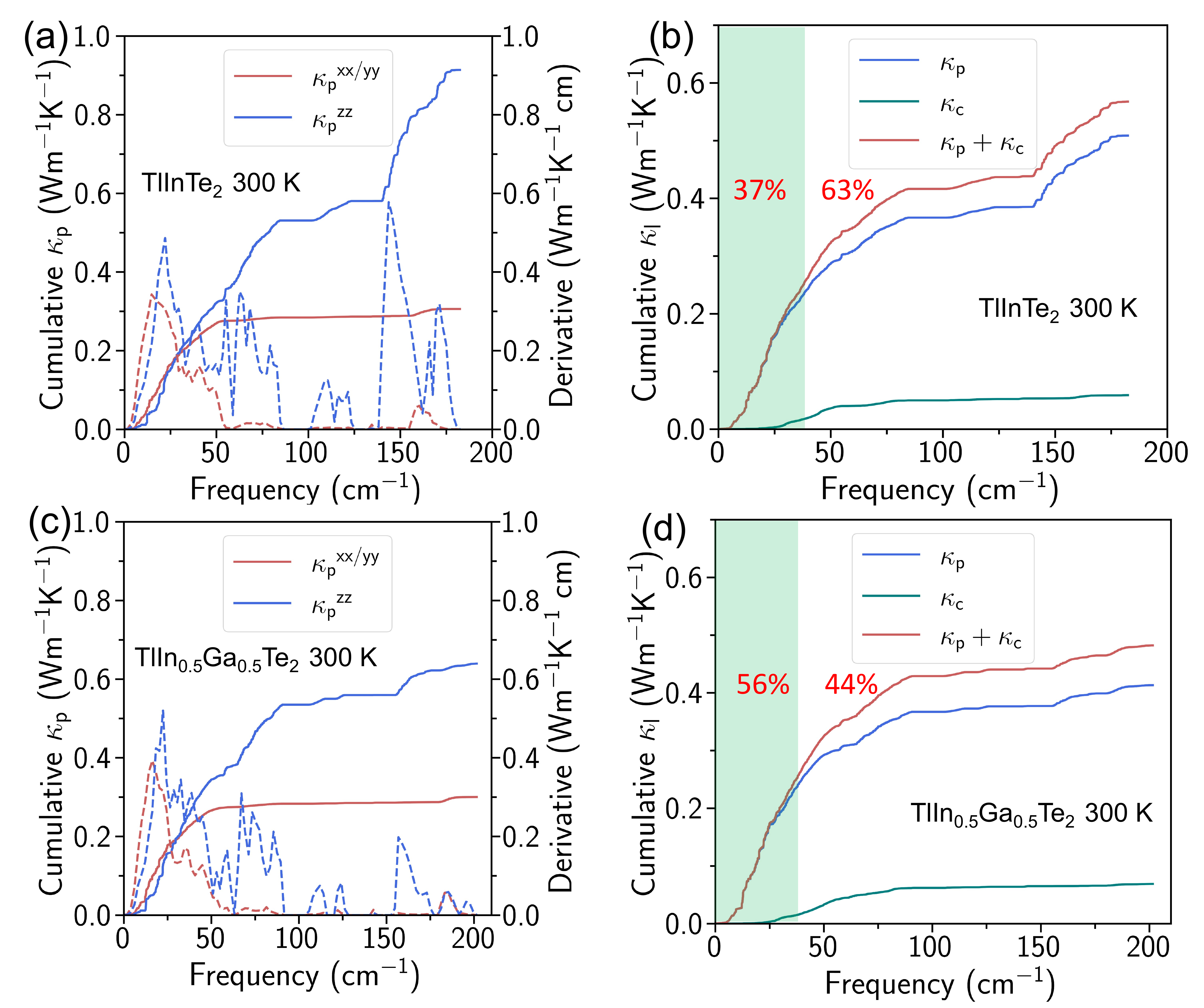}
  \caption{Cumulative $\kappa_p$ as a function of phonon frequency and its derivative of (a) TlInTe$_2$ and (c) TlIn$_{0.5}$Ga$_{0.5}$Te$_2$ along $xx/yy$ and $zz$-direction at 300 K. Average cumulative $\kappa_p$, $\kappa_c$, and $\kappa_l=\kappa_p+\kappa_c$ of (b) TlInTe$_2$ and (d) TlIn$_{0.5}$Ga$_{0.5}$Te$_2$ at 300 K. Green and white regions of (c) and (d) indicate acoustic and optical regions, respectively.}
  \label{fig:cum_kl}
\end{figure}

FIG.~\ref{fig:kl}a represents the temperature-dependent average population conductivity ($\kappa_p$) arising from particle-like phonon propagation, coherence conductivity ($\kappa_c$) due to wave-like phonon tunnelling, and total lattice thermal conductivity ($\kappa_l=\kappa_p+\kappa_c$) of TlInTe$_2$, calculated using LWTE and compared with the available experimental $\kappa_l$ data. $\kappa_p$ arises from the diagonal elements of the heat-flux operator, which represents heat transport mediated by the propagation of phonon modes. $\kappa_p$ decreases with temperature with 1/T, following the BTE model due to the enhanced phonon-phonon scattering at the higher temperatures. On the other hand, $\kappa_c$ comes from the off-diagonal elements of the heat-flux operator, which describes the wave-like tunnelling of thermal energy due to the coupling between quasi-degenerate phonon modes. With increasing temperature, enhanced phonon–phonon scattering leads to a broadening of phonon linewidths, which creates stronger coupling between quasi-degenerate phonon modes. As a result, $\kappa_c$ exhibits a slight increase with temperature. In TlInTe$_2$ $\kappa_p$ decreases from 0.509 Wm$^{-1}$K$^{-1}$ at 300 K to 0.193 Wm$^{-1}$K$^{-1}$ at 800 K, whereas $\kappa_c$ shows a slight increases from 0.059 Wm$^{-1}$K$^{-1}$ to 0.070 Wm$^{-1}$K$^{-1}$ over the same temperature range. Thus, the $\kappa_l$ of TlInTe$_2$ is predominantly governed by $\kappa_p$, and the contribution of $\kappa_c$ is negligible. The total $\kappa_l$ decreases from 0.568 Wm$^{-1}$K$^{-1}$ at 300 K to 0.263 Wm$^{-1}$K$^{-1}$ at 800 K. The calculated $\kappa_l$ values are in good agreement with previously reported experimental data by Matsumoto et al.\cite{matsumoto2008systematic} and Jana et al.,\cite{jana2017intrinsic}, which validates the reliability of our present theoretical calculations (FIG. \ref{fig:kl}a). Further we applied boundary scattering, which scatters the long-wavelength phonons and plays a significant role in limiting the $\kappa_l$ at the lower temperatures. With boundary scattering of 30 nm grain boundary size, the calculated $\kappa_l$ reproduces the experimental data of Jana et al.\cite{jana2017intrinsic} more accurately at low temperatures (FIG. \textcolor{blue}{S3, Supplemental Material}). Pal et al.\cite{pal2021microscopic} showed that $\kappa_l$ using LWTE coupled with phonon frequency renormalization at finite temperatures using the self-consistent phonon (SCPH) method\cite{tadano2015self} also reproduces the experimental data of Matsumoto et al.\cite{matsumoto2008systematic} The low $\kappa_l$ in TlInTe$_2$ is attributed to the bonding heterogeneity and rattling of Tl atoms, which scatters the heat-carrying acoustic phonons significantly and reduce the sound velocity of the system.

When all the In atoms are replaced by Ga atoms, the $\kappa_l$ of TlGaTe$_2$ has increased to 0.736 Wm$^{-1}$K$^{-1}$ at 300 K due to lighter Ga atoms compared to In atoms (FIG.~\ref{fig:kl}b). The calculated $\kappa_l$ of TlGaTe$_2$ is also in good agreement with available experimental data (FIG.~\textcolor{blue}{S4, Supplemental Material}). However, when 50\% of In atoms of TlInTe$_2$ is replaced by Ga atoms, the $\kappa_l$ of TlIn$_{0.5}$Ga$_{0.5}$Te$_2$ is decreased to 0.482 Wm$^{-1}$K$^{-1}$ at 300 K (FIG.~\ref{fig:kl}b). Despite the lighter mass of Ga, this reduction indicates that the decrease in $\kappa_l$ is not governed by mass effects, but is instead driven by modifications in phonon dispersion. FIG.~\textcolor{blue}{S5, Supplemental Material} shows the components of $\kappa_p$ and $\kappa_c$ along $xx/yy$ and $zz$-directions of TlIn$_{0.5}$Ga$_{0.5}$Te$_2$. It is observed that both $\kappa_p$ and $\kappa_c$ are higher along the $z$-direction, with the anisotropy being more pronounced for $\kappa_p$. Thus the $\kappa_l$ is significantly higher along $z$-direction (0.745 Wm$^{-1}$K$^{-1}$ at 300 K) than $xx/yy$-directions (0.351 Wm$^{-1}$K$^{-1}$ at 300 K). This anisotropic behavior originates from the one-dimensional tetrahedral chain along the crystallographic $c$-axis, which allows better heat transport along $zz$-direction, while the weaker interchain interactions limit heat conduction in the $ab$-plane.

\begin{figure}[!t] 
  \centering
  \includegraphics[width=0.9\columnwidth]{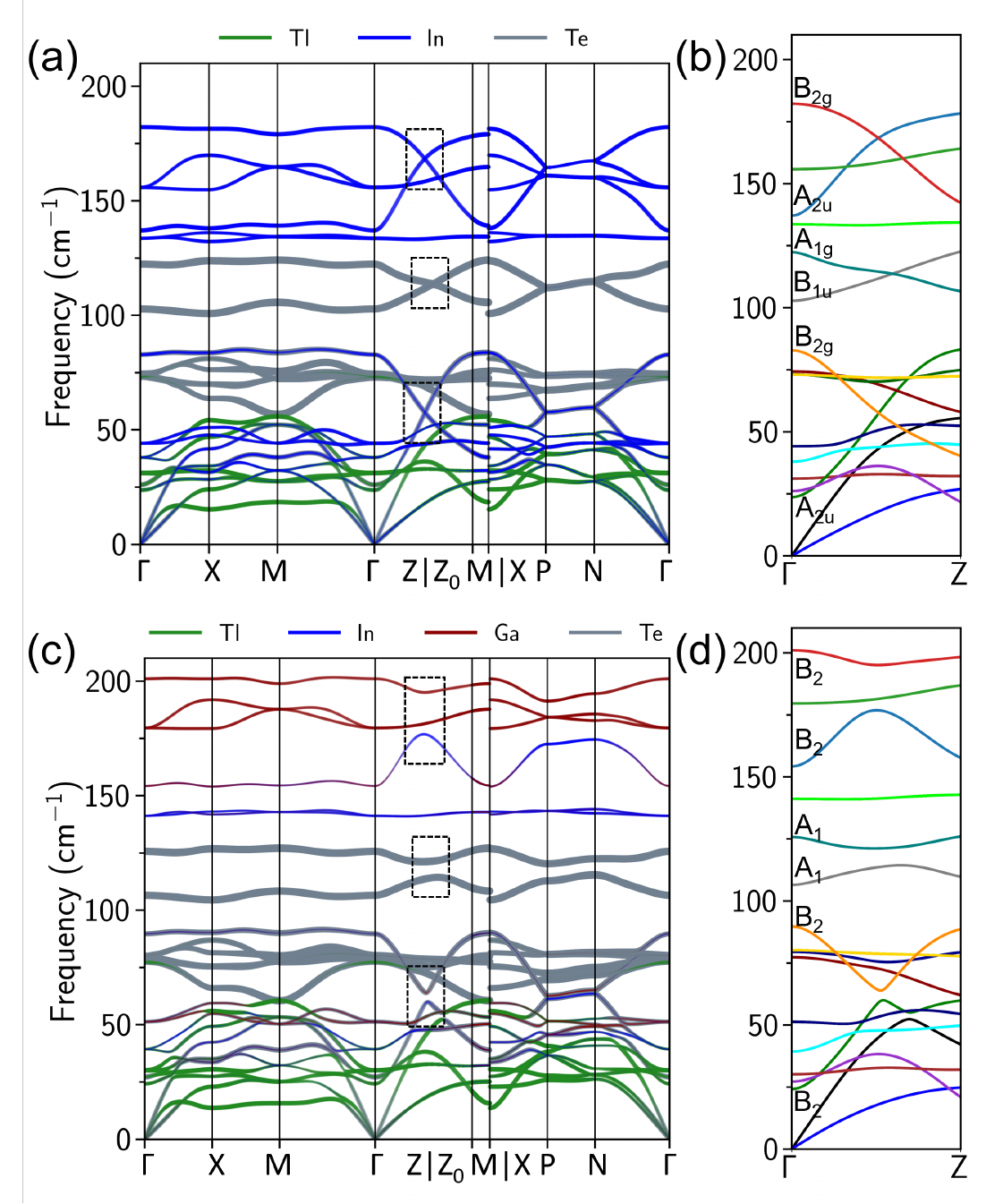}
  \caption{Atom-projected phonon dispersion of (a) TlInTe$_2$ and (c) TlIn$_{0.5}$Ga$_{0.5}$Te$_2$. The phonon crossing points in TlInTe$_2$, which which evolve into avoided crossing in TlIn$_{0.5}$Ga$_{0.5}$Te$_2$ along $\Gamma$--Z direction, are highlighted by dotted black boxes. Enlarged view of the $\Gamma$--Z direction of the phonon dispersion are shown in (b) and (d) for TlInTe$_2$ and TlIn$_{0.5}$Ga$_{0.5}$Te$_2$, respectively. The irreducible representations of the phonon modes at the $\Gamma$ point corresponding to the crossing and avoided-crossing branches are also indicated.}
  \label{fig:phonon}
\end{figure}

To understand the specific phonon contribution in the $\kappa_l$, we have plotted cumulative $\kappa_l$ as a function of frequency of TlInTe$_2$ and TlIn$_{0.5}$Ga$_{0.5}$Te$_2$ at 300 K, as shown in FIG.~\ref{fig:cum_kl}.  The cumulative $\kappa_p$ with frequency along $xx/yy$ and  $zz$-directions of TlInTe$_2$ shows that along the $xx/yy$-direction maximum amount of $\kappa_p$ is contributed by the acoustic region ($\sim$61\%), whereas the contribution of optical region in less ($\sim$29\%). However, along the $zz$-direction, an opposite trend is observed. In this direction, only $\sim$23\% of $\kappa_p$ is originates from the acoustic region and $\sim$77\% of $\kappa_p$ is contributed by optical region (FIG.~\ref{fig:cum_kl}a). After adding $\kappa_c$ with $\kappa_p$, the total $\kappa_l$ comprises approximately 37\% acoustic and 63\% optical contributions (Green and white regions is FIG.~\ref{fig:cum_kl}b indicates acoustic and optical regions, respectively). Interestingly, along the $zz$-direction, we observe a sharp jump in cumulative $\kappa_p$ at $\sim$150 cm$^{-1}$, which indicates the optical phonons around this frequency range plays an important contribution in the $\kappa_l$ (FIG.~\ref{fig:cum_kl}a). This is further highlighted by a sharp peak in the derivative of the cumulative $\kappa_p$ along $zz$-direction at $\sim$150 cm$^{-1}$. We observe that this sharp peak is higher than the peak in the acoustic region (FIG.~\ref{fig:cum_kl}a). These results confirm that the thermal transport of TlInTe$_2$ is dominated by optical phonons. In TlGaTe$_2$, similar optical phonon dominated $\kappa_l$ nature is observed, as shown in FIG.~\textcolor{blue}{S7, Supplemental Material}.

In TlIn$_{0.5}$Ga$_{0.5}$Te$_2$, the cumulative $\kappa_p$ with frequency shows similar trend along the $xx/yy$-direction, however a pronounced suppression is observed along the $zz$-direction (FIG.~\ref{fig:cum_kl}c). In particular, the sharp peak in the derivative of cumulative $\kappa_p$ at $\sim$150 cm$^{-1}$ in TlInTe$_2$ shifts to $\sim$160 cm$^{-1}$ with a significantly reduced value in TlIn$_{0.5}$Ga$_{0.5}$Te$_2$. Consequently, the optical phonon contribution in $\kappa_p$ along the $zz$-direction is reduced to $\sim$64\% in TlIn$_{0.5}$Ga$_{0.5}$Te$_2$ from $\sim$77\% in TlInTe$_2$. Thus, the overall optical phonon contribution in $\kappa_l$ is reduced to $\sim$44\% (white region of FIG.~\ref{fig:cum_kl}d). The suppression of optical phonon contribution is further evident from the absolute values. At 300 K, the acoustic and optical contributions in $\kappa_l$ for TlInTe$_2$ are 0.210 Wm$^{-1}$K$^{-1}$ and 0.358 Wm$^{-1}$K$^{-1}$, respectively, whereas in TlIn$_{0.5}$Ga$_{0.5}$Te$_2$, those values are 0.269 Wm$^{-1}$K$^{-1}$ and 0.213 Wm$^{-1}$K$^{-1}$. Thus, these results clearly demonstrates that optical phonons are responsible in suppressing the $\kappa_l$ in the partial Ga-substituted TlInTe$_2$.

To elucidate the role of specific optical phonon modes in lowering the $\kappa_l$ of TlIn$_{0.5}$Ga$_{0.5}$Te$_2$, we have calculated atom projected phonon dispersion of both compounds, as shown in FIG.~\ref{fig:phonon}a and \ref{fig:phonon}c. FIG.~\ref{fig:phonon}a shows that in TlInTe$_2$ the low frequency optical modes are dominated by Tl atoms, whereas mid and high frequency optical modes are contributed by In and Te atoms. On a closer inspection of the phonon dispersion of TlInTe$_2$, few interesting features are observed. We find several highly dispersive optical phonon modes along the $\Gamma$--Z, X--P and N--$\Gamma$ directions. These highly dispersive modes are mainly associated with In atoms, while the Te-dominated modes are comparatively less dispersive. Interestingly, several phonon branch crossings between the optical phonon branches are observed along the $\Gamma$--Z direction, highlighted by dotted black boxes in Fig.~\ref{fig:phonon}a. To examine these crossings more clearly, an enlarged view of the $\Gamma$--Z direction is shown in Fig.~\ref{fig:phonon}b. In addition, phonon branch connectivity was analyzed using the eigenvector overlap, implemented in phonopy code,\cite{togo2023first, togo2023implementation} which is defined as 
\begin{equation}
S_{\lambda\lambda'}(q,q+\Delta q)
=
\left|
\braket{e_{\lambda}(q)\,|\,e_{\lambda'}(q+\Delta q)}
\right|,
\label{eq:overlap}
\end{equation}
where $e_{\lambda}(q)$ is phonon eigen vector at $\lambda$ mode at $q$ point. Using this equation, the overlap between eigenvectors of two adjacent points are calculated and the eigenvectors with highest overlap are considered most similar, so they are assumed to belong to same phonon branch. We have used different colours to separate different phonon branches. FIG.~\ref{fig:phonon}b shows that along $\Gamma$--Z direction, the bands corresponds to the black dot box in FIG. \ref{fig:phonon}a are clearly crossing. We also calculated irreducible representation at the $\Gamma$ point of the phonon branches participated in crossing each other. The irreducible representation shows that the phonon branches belongs to different symmetry representations ((A$_{\mathrm{2u}}$, B$_{\mathrm{2g}}$), (B$_{\mathrm{1u}}$, A$_{\mathrm{1g}}$), and (A$_{\mathrm{2u}}$, B$_{\mathrm{2g}}$)) are crossing, which proves that phonon modes belonging to different symmetry groups are noninteracting and they cross each other without hybridization. TlGaTe$_2$ also exhibits similar phonon branch crossing along $\Gamma$--Z point, as shown in FIG.~\textcolor{blue}{S8, Supplemental Material}.    

\begin{figure}[!t] 
  \centering
  \includegraphics[width=0.8\columnwidth]{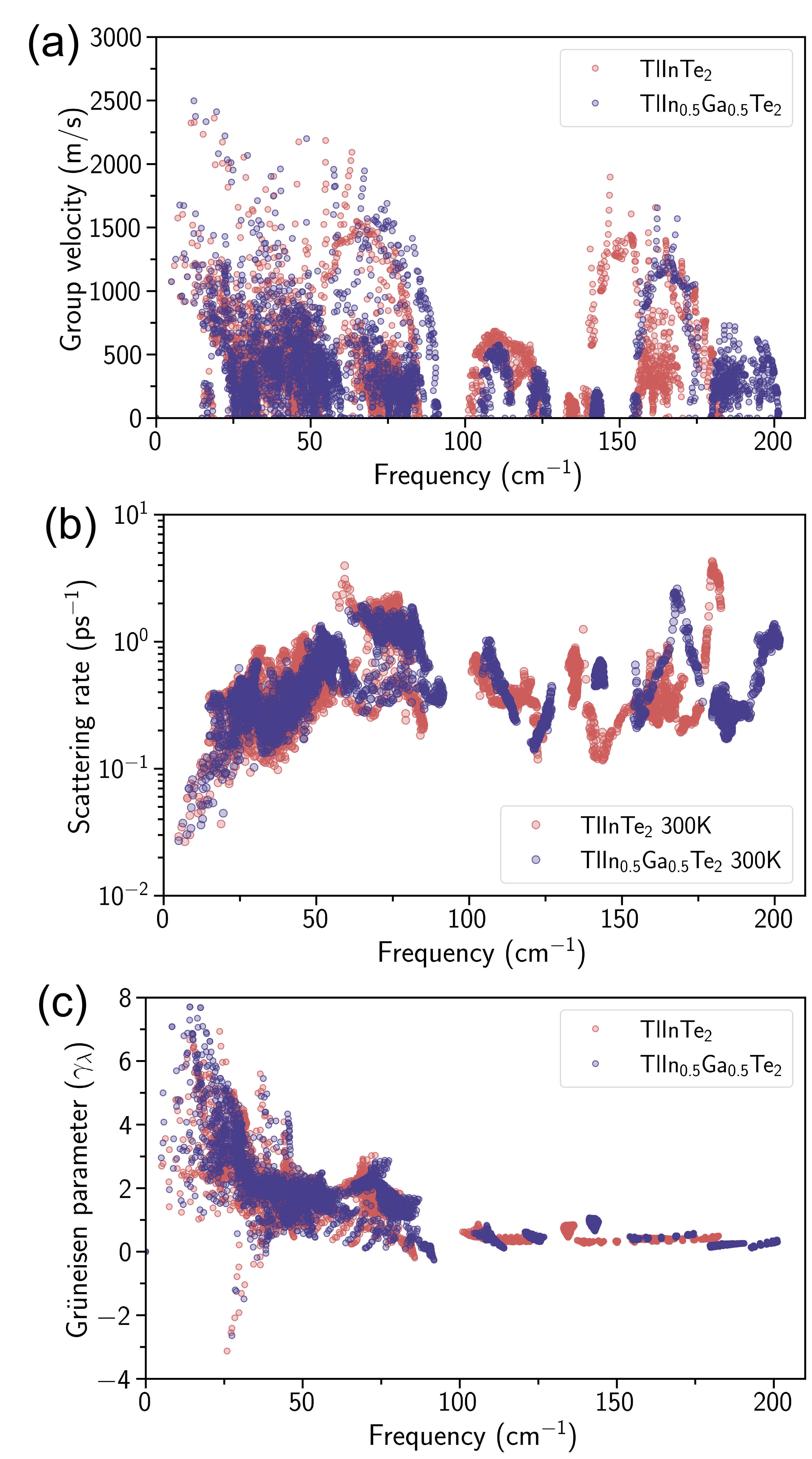}
  \caption{Comparison of frequency-dependent (a) phonon group velocity ($v_g$), (b) scattering rate, and (c) Gr\"uneisen parameter ($\gamma_{\lambda}$) of TlInTe$_2$ and TlIn$_{0.5}$Ga$_{0.5}$Te$_2$.}
  \label{fig:gv}
\end{figure}

The atom-projected phonon dispersion of TlIn$_{0.5}$Ga$_{0.5}$Te$_2$ (FIG.~\ref{fig:phonon}c) shows an increased maximum phonon frequency compared to TlInTe$_2$ and high-frequency range is mainly dominated by Ga due to its lower mass than In atoms. Surprisingly, the phonon branches that intersect in TlInTe$_2$ evolve into avoided crossings in TlIn$_{0.5}$Ga$_{0.5}$Te$_2$ as highlighted by the dotted black boxes in FIG.~\ref{fig:phonon}c. The enlarged view of phonon dispersion along $\Gamma$--Z direction in FIG.~\ref{fig:phonon}d clearly shows the repulsion between the interacting phonon branches. Additionally, the irreducible representation at $\Gamma$ point shows same symmetry corresponding to the interacting branches, which are (B$_{\mathrm{2}}$,B$_{\mathrm{2}}$), (A$_{\mathrm{1}}$,A$_{\mathrm{1}}$), and (B$_{\mathrm{2}}$,B$_{\mathrm{2}}$). This symmetry modification in the phonon modes is associated with the partial Ga doping in pristine TlInTe$_2$. Thus, the phonon modes with same symmetry no longer intersect; instead, they hybridize and repel each other, giving rise to characteristic gap opening at the crossing points. 
~
\begin{table}[t]
    \caption{Mode-average phonon group velocity ($\bar v$), relaxation time ($\bar{\tau}$) at 300 K, and Gr\"uneisen parameter ($\bar{\gamma}$) of TlInTe$_2$ and TlIn$_{0.5}$Ga$_{0.5}$Te$_2$. }
    \label{tab:compare}
    \centering
    \setlength{\tabcolsep}{12pt}
    \begin{tabular} {ccc}
         \hline \hline
          Parameter & TlInTe$_2$ & TlIn$_{0.5}$Ga$_{0.5}$Te$_2$ \\
        
           \hline
           $\bar v_g$ (m/s) & 704.20 & 638.56 \\
           
           $\bar{\tau}$ (ps) & 4.04 & 3.90 \\

           $\bar{\gamma}$ & 1.56 & 1.59 \\
           \hline \hline
    \end{tabular}
\end{table}

The emergence of avoided crossings among optical phonon branches leads to a flattening of the optical phonon dispersion, resulting in a reduction of the corresponding phonon group velocity ($v_g$). As shown in Fig.~\ref{fig:gv}(a), the optical phonon $v_g$ in TlIn$_{0.5}$Ga$_{0.5}$Te$_2$ is noticeably suppressed compared to TlInTe$_2$, particularly within the frequency ranges 100-125 cm$^{-1}$ and 150-200 cm$^{-1}$, which coincide with the avoided-crossing regions. The phonon-phonon scattering rate and Gr\"uneisen parameter ($\gamma_{\lambda}$) vs frequency plot for TlInTe$_2$ and TlIn$_{0.5}$Ga$_{0.5}$Te$_2$ are shown in FIG.~\ref{fig:gv}b and \ref{fig:gv}c. For better comparison, we have calculated mode average $v_g$, lifetime ($\tau$), and $\gamma$ for both compounds using the equations:\cite{li2021optical}
\begin{equation}
    \bar{v_g} = \sqrt{\frac{\sum_{\lambda} C_{V \lambda} v_{\lambda}^{2} \tau_{\lambda}}{\sum_{\lambda} C_{V \lambda} \tau_{\lambda}}},
\end{equation}
\begin{equation}
    \bar{\tau} = \frac{\sum_{\lambda} C_{V \lambda} v_{\lambda}^{2} \tau_{\lambda}}{\sum_{\lambda} C_{V \lambda}  v_{\lambda}^{2}},
\end{equation}
\begin{equation}
    \bar{\gamma} = \frac{\sum_{\lambda} \left | \gamma_{\lambda} \right |C_{V \lambda}} {\sum_{\lambda} C_{V \lambda}},
\end{equation}
where $C_V$ is phonon specific heat and $\lambda$ is phonon mode. The calculated values are summarized in Table~\ref{tab:compare}. We observe that $\bar v_g$ is strongly reduced in TlIn$_{0.5}$Ga$_{0.5}$Te$_2$, which is associated with the emergence of avoided crossings between the optical phonon branches. The higher $\bar{\gamma}$ also increases phonon-scattering rate slightly in TlIn$_{0.5}$Ga$_{0.5}$Te$_2$ and reduces $\bar{\tau}$. As $\kappa_p =\sum_{\lambda}C_{V\lambda}v^2_{\lambda}\tau_{\lambda}$ within the phonon gas model under the single-mode approximation (SMA) method, the larger reduction in $\bar v_g$ indicates that the lower $\kappa_l$ in TlIn$_{0.5}$Ga$_{0.5}$Te$_2$ is primarily governed by reduced $v_g$, while enhanced anharmonic scattering provides an additional contribution in suppressing $\kappa_l$. 

\begin{figure}[!h] 
  \centering
  \includegraphics[width=1\columnwidth]{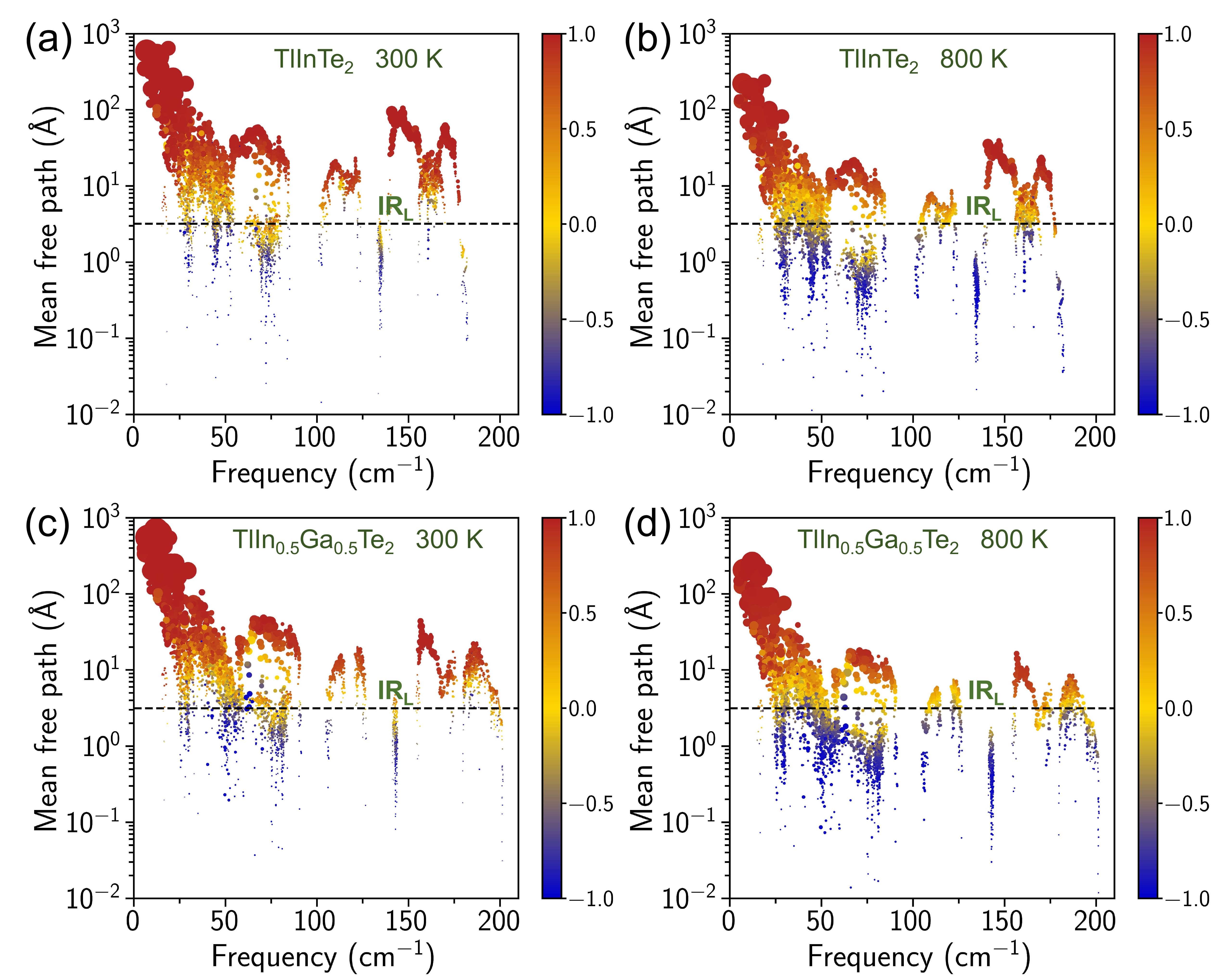}
  \caption{Phonon mean free path ($\Lambda$) as a function of phonon frequency of TlInTe$_2$ at (a) 300 K, and (b) 800 K. $\Lambda$ as a function of phonon frequency of TlIn$_{0.5}$Ga$_{0.5}$Te$_2$ at (c) 300 K, and (d) 800 K.}
  \label{fig:mean_free}
\end{figure}

Further, we have compared the mode-resolved phonon mean free path ($\Lambda$) of TlInTe$_2$ and TlIn$_{0.5}$Ga$_{0.5}$Te$_2$ at 300 K and 800 K, as shown FIG.~\ref{fig:mean_free}. Here, the area of each circle is proportional to the ratio of thermal conductivity of that mode ($\kappa_p(\mathbf q)_s+\kappa_c(\mathbf q)_s$) to total lattice thermal conductivity ($\kappa_l$) of the system. Thus, the size of each circle represents the relative contribution of an individual phonon mode to heat transport. The colour of each circle represents the quantity $c=\frac{\kappa_p(\mathbf q)_s-\kappa_c(\mathbf q)_s}{\kappa_p(\mathbf q)_s+\kappa_c(\mathbf q)_s}$, which characterizes the degree of particle-like and wave-like nature of a phonon mode. So, the brown region ($c=+1$) corresponds to purely particle-like phonon transport, whereas the blue region ($c=-1$) indicates wave-like phonon coherence. The intermediate yellow region represents the coexistence of both particle-like and wave-like transport characteristics. The dashed line represents the ``Ioffe-Regel limit" ($\mathrm{IR_L}$), which sets a threshold limit so that phonon modes above $\mathrm{IR_L}$ are considered as propagating modes, whereas modes below it are regarded as non-propagating or localized vibrations.\cite{agne2018minimum, luo2020vibrational, taraskin2000ioffe} $\mathrm{IR_L}$ is approximated as average interatomic distance of a system.\cite{simoncelli2022wigner} For TlInTe$_2$ and TlIn$_{0.5}$Ga$_{0.5}$Te$_2$, the calculated $\mathrm{IR_L}$ are 3.19 \AA\ and 3.14 \AA, respectively. At 300 K, most of the phonon modes are found above $\mathrm{IR_L}$ and participated in $\kappa_p$ in both systems. With increasing temperature 300 K to 800 K, $\Lambda$ decrease significantly due to higher phonon scattering at higher temperature, causing a larger number of phonon modes to cross the $\mathrm{IR_L}$, and consequently the contribution of wave-like phonon coherence increases. More importantly, we observe that the $\Lambda$ of phonon modes between 150--200 cm$^{-1}$ in TlIn$_{0.5}$Ga$_{0.5}$Te$_2$ is lower than that those of TlInTe$_2$. In addition, the relative thermal conductivity contribution of these modes, represented by the circle size, is significantly suppressed in TlIn$_{0.5}$Ga$_{0.5}$Te$_2$ compared to the acoustic phonon modes. This results further confirm that avoided crossings at the optical phonon suppress the phonon transport and leads to reduce $\kappa_l$ in TlIn$_{0.5}$Ga$_{0.5}$Te$_2$. 

\section{Conclusion}
In this work, we demonstrated that optical–optical avoided crossings provide an effective mechanism for suppressing lattice thermal conductivity in TlIn$_{0.5}$Ga$_{0.5}$Te$_2$. Using the Wigner formalism of thermal transport (LWTE), we observe populations conductivity ($\kappa_p$) is the primary transport channel in these compounds. In pristine TlInTe$_2$, the calculated $\kappa_l$ is 0.568 Wm$^{-1}$K$^{-1}$ at 300 K, where optical phonons dominate the heat transport and contribute nearly 63\% of $\kappa_l$. Upon 50\% Ga substitution in TlInTe$_2$, the $\kappa_l$ of TlIn$_{0.5}$Ga$_{0.5}$Te$_2$ is reduced to 0.482 Wm$^{-1}$K$^{-1}$ at 300 K. This reduction in $\kappa_l$ primarily comes from the optical phonon region, as shown in the cumulative $\kappa_l$ vs frequency analysis. The phonon dispersion of TlInTe$_2$ exhibits several phonon crossing points in the optical region, which evolve into avoided crossings in TlIn$_{0.5}$Ga$_{0.5}$Te$_2$. The irreducible representation analysis at $\Gamma$-point shows that phonon branches belongs to different symmetry representations in TlInTe$_2$ are crossing, which modified into same symmetry representation in TlIn$_{0.5}$Ga$_{0.5}$Te$_2$. The phonon modes with same symmetry do not intersect; instead they repel each other and giving rise to characteristic gap openings at the crossing points. The avoided crossings in TlIn$_{0.5}$Ga$_{0.5}$Te$_2$ reduce phonon group velocity strongly within the frequency ranges 100--125 cm$^{-1}$ and 150--200 cm$^{-1}$. and consequently the optical phonon contribution to $\kappa_p$ is reduced to 44 \%. Further analysis of the mode-averaged transport parameters, $\bar v_g$, $\bar{\tau}$ and $\bar{\gamma}$, reveals a significant suppression of $\bar v_g$ in TlIn$_{0.5}$Ga$_{0.5}$Te$_2$, indicating that the reduction in $\kappa_l$ is primarily governed by reduced phonon group velocity induced by avoided crossings at the optical region.  The slightly higher $\bar{\gamma}$ and lower $\bar{\tau}$ indicates that the enhanced anharmonic scattering provides an additional contribution in lowering $\kappa_l$. Thus, our work establishes that symmetry-modified optical-optical avoided crossings can serve as an effective mechanism for reducing $\kappa_l$ in compounds where optical phonons significantly contribute to heat transport.

\begin{acknowledgements}
S.P. thanks CSIR, Government of India, for providing the SRF CSIR Fellowship. S.K.P. acknowledges ANRF J. C. Bose National Fellowship (File number: ANRF/SKP/4719) and ANRF, Govt. of India for funding. S.P. and S.K.P. thanks JNCASR and Param Yukti, under the National Supercomputing Mission (NSM), Government of India for providing computation facilities.   
\end{acknowledgements}

\bibliographystyle{apsrev4-2}
\bibliography{mybib}
\end{document}